\documentclass{article}

\usepackage{arxiv}

\usepackage[utf8]{inputenc} 
\usepackage[T1]{fontenc}    
\usepackage{hyperref}       
\usepackage{url}            
\usepackage{booktabs}       
\usepackage{amsfonts}       
\usepackage{nicefrac}       
\usepackage{microtype}      
\usepackage{lipsum}		

\usepackage{booktabs} 
\usepackage{color}
\usepackage[ruled]{algorithm2e} 
\usepackage{float}

\newcommand{\es}{\enspace}

\newcommand{\mbf}[1]{\mathbf{#1}}

\usepackage{hyperref}
\usepackage[export]{adjustbox}
\usepackage{listings}
\usepackage{ulem}
\usepackage{float}
\usepackage{amsmath,amssymb}

\title{SODECL: An Open Source Library for Calculating Multiple Orbits of a System of Stochastic Differential Equations in Parallel} 
\date{February 17, 2020}	

\author{
	Eleftherios Avramidis \\
	Research Computing Services\\
	University of Cambridge\\
	Cambridge, UK\\
	\texttt{ea461@cam.ac.uk} \\
	\And
	Marta Lalik \\
	University of Exeter, now at Isomerase Therapeutics Ltd. \\
	Cambridge, UK\\
	\texttt{marta\_lalik@hotmail.com} \\
	\And
	Ozgur E. Akman \\
	College of Engineering,	Mathematics \& Physical Sciences \\
	University of Exeter \\
	Exeter, UK\\
	\texttt{O.E.Akman@exeter.ac.uk} \\
}

\begin{document}
\maketitle
	
\begin{abstract}

Stochastic differential equations (SDEs) are widely used to model systems affected by random processes. In general, the analysis of an SDE model requires numerical solutions to be generated many times over multiple parameter combinations. However, this process often requires considerable computational resources to be practicable. Due to the embarrassingly parallel nature of the task, devices such as multi-core processors and graphics processing units (GPUs) can be employed for acceleration.

Here, we present {\bf SODECL} (\url{https://github.com/avramidis/sodecl}), a software library that utilises such devices to calculate multiple orbits of an SDE model. To evaluate the acceleration provided by SODECL, we compared the time required to calculate multiple orbits of an exemplar stochastic model when one CPU core is used, to the time required when using all CPU cores or a GPU. In addition, to assess scalability, we investigated how the model size affected execution time on different parallel compute devices.

Our results show that when using all 32 CPU cores of a high-end high-performance computing node, the task is accelerated by a factor of up to $\simeq$6.7, compared to when using a single CPU core. Executing the task on a high-end GPU yielded accelerations of up to $\simeq$4.5, compared to a single CPU core.

\end{abstract}

%
%


\maketitle



\section{Introduction}

\subsection{Modelling dynamic systems using SDEs}

Noise affects the behaviour of a vast number of physical and biological phenomena, such as gene expression~\cite{Kaern2005,Keren2015}, the transmission of nerve impulses~\cite{Faisal2008,Deco2009} and the dynamics of electronic circuits~\cite{Darabi2000,Scholten2003,Kaern2005}. Systems that are affected by noise are commonly modelled using sets of stochastic differential equations (SDEs)~\cite{Allen2008}. In such models, each equation represents the rate of change of a system variable ({\it e.g.} voltage) with time and can be dependent on one or more parameters ({\it e.g.} capacitance). Moreover, one or more equations are also affected by a noise process. The It$\hat{\rm o}$ form of a first-order coupled SDE system is
\begin{equation}
dX_i(t) = f_i(t, \mathbf{X}(t),\mathbf{p})\, dt+ \sum_{j=1}^{m}g_{ij}(t,\mathbf{X}(t),\mathbf{p})\, dW_j(s), 
\label{eq:SODEs}
\end{equation}
where $t\geq t_0$ is time, $\mathbf{X}=\left(X_1,\ldots,X_n\right)$ is the state vector, $\mathbf{X}_0=\mathbf{X}(t_0)$ specifies the initial state of the system, $\mathbf{p}=\left(p_1,\ldots,p_d\right)$ is a vector of parameter values, $\{W_1(t),\ldots,W_m(t)\}$ are independent, scalar Wiener processes, and the functions $\{f_i(t,\mathbf{X},\mathbf{p}): 1\leq i \leq n\}$ and $\{g_{ij}(t,\mathbf{X},\mathbf{p}):1\leq i \leq n, 1\leq j \leq m\}$ are the drift and diffusion coefficients, respectively \cite{Platen1999,Higham2001,Kloeden2011}. Setting the diffusion coefficients $g_{ij}$ to 0 reduces the system to a set of deterministic, ordinary differential equations (ODEs).

\subsection{Optimising the parameters of SDE models}

By varying the parameters $\mbf{p}$ of the SDE system~(\ref{eq:SODEs}), different classes of dynamical behaviour can be observed ({\it e.g.} transitions between bursting and spiking in neural models). However, in many cases, the parameters are difficult to measure experimentally. Consequently, optimisation methods are often employed to find the particular parameter combinations that most closely reproduce the experimentally-measured behaviour of the system of interest~\cite{Palancz2016}. Performing this parameter optimisation step in a robust, automated fashion is a critical step in the construction and analysis of biological models~\cite{Akman08,Akman10,Lillacci2010,Akman12,Avramidis2017,Doherty17}, because determining the optimal parameter values enables alternative models to be systematically ranked and experimentally testable predictions to be formulated~\cite{Cullen1996,Johnson2004,Ashyraliyev2009,Cedersund2009,Lillacci2010,Slezak2010,Sun2012,Avramidis2017}. The assumptions made in constructing a given model can then be rigorously assessed, and insights obtained into how the model could be modified so as to improve the accuracy of its predictions~\cite{Ashyraliyev2009,Cedersund2009,Lillacci2010,Slezak2010,Avramidis2017}.

However, optimising the parameters $\mbf{p}$ of system (\ref{eq:SODEs}) can incur a high computational cost, depending on the forms of $f_i$ and $g_{ij}$, the system size $n$, and the number of noise terms $m$. Large systems (high $n$ values), multiple noise processes (high $m$ values) and computationally demanding drift and diffusions terms ({\it e.g.} terms containing transcendental functions) can all increase the computational cost considerably. Furthermore, an additional cost is associated with the generation of the random numbers required to simulate the noise processes. The computational cost can also be significantly increased by the particular parameter optimisation method employed; for example, when using a population-based method -- such as an evolutionary algorithm (EA) or a particle swarm optimiser (PSO) -- the equations have to be integrated multiple times over different parameter values to explore the underlying fitness landscape~\cite{He2001,Witt2008,Cedersund2016}. Additionally, a resampling approach might be required to mitigate the effects of noise and uncertainty in model evaluation, thereby increasing the computational load further~\cite{Fieldsend2015,Doherty17}.

In general, the evaluation of an SDE model for the optimisation methods mentioned above involves individual tasks with no interdependence or communication. These tasks can therefore be parallelised in a straightforward manner, and hence the optimisation problem is referred to as embarrassingly or pleasingly parallel~\cite{Navarro2014}.

\subsection{Accelerating the optimisation process using multi-core and many-core devices}

The significant computational demands of such embarrassingly parallel tasks means that high-performance computing (HPC) clusters ({\it i.e.} multiple connected computers) are required to obtain results within a reasonable time frame. Recently, competitively-priced CPU models that provide between 6 and 16 physical cores have become commercially available. In the case of workstation CPUs, there are now models with 32 cores.

Moreover, current GPU models, which are commonly used for graphics generation in personal computers and workstations, can also be leveraged for scientific computing~\cite{Papadrakakis2011,Mantas2016}. They contain a large number of processing units -- each of which is relatively slow compared to the cores of a CPU -- and are optimised for single instruction, multiple data (SIMD) processing. This makes them well-suited to large-scale parallelisation. Indeed, the collective processing power of a GPU can exceed that of a single CPU for certain tasks. Moreover, depending on the motherboard model, multiple GPUs can be installed on a single desktop/workstation computer, boosting the potential processing power even further.

In a previous study, we demonstrated that a desktop computer with a GPU was able to robustly optimise the parameters of a spiking neuron model to experimental eye movement data using a multi-objective EA~\cite{Avramidis2017}. As part of this work, we showed that the most computationally intensive part of the optimisation process was the numerical integration of the model over the multiple parameter combinations comprising the EA population at each generation~\cite{Avramidis2017}. This task can be executed in an embarrassingly parallel fashion, enabling the current trend for heterogenous HPC architectures to be exploited. Accordingly, in our spiking neuron model, numerical integration was executed on the GPU, while all the EA operations (mutation, crossover, fitness evaluation {\it etc.}) were executed on the CPU~\cite{Avramidis2017}. Utilising the GPU in this manner yielded a speedup of up to $\approx$20, compared with a high-end CPU.

\subsection{CUDA and OpenCL}

Programs that are executed on a GPU have to be written using a specific programming framework. The two most commonly used frameworks are CUDA (Compute Unified Device Architecture) and OpenCL (Open Computing Language)~\cite{Demidov2013}. The CUDA framework is a proprietary architecture specifically designed to be run on NVIDIA compute devices. By contrast, OpenCL is a royalty-free standard for general purpose parallel programming across CPUs and GPUs, giving software developers portable and efficient access to the power of these heterogeneous processing platforms~\cite{OpenCL2.1}. OpenCL includes a cross-platform intermediate language for writing functions (kernels), which are executed on OpenCL-supported devices, together with an application programming interface (API) that is used to coordinate the parallel computations across these devices. A simple OpenCL program involves the following steps: identifying the OpenCL device; compiling the OpenCL code that will be run on the device; copying the data to the device; performing the computation; copying the results back from the device.

A software library equipped with OpenCL functionality can use CPUs and GPUs, removing the need to develop different libraries for each processor type.

\subsection{A new library for numerically integrating SDE models on OpenCL-supported devices}

Here, we present SODECL, a C++ library that uses OpenCL to calculate multiple orbits of an SDE (or ODE) system in an embarrassingly parallel way. We focus on presenting the library for solving SDE models, for which orbits are calculated using the Euler-Maruyama method~\cite{Saito1996,Higham2001}. ODE orbits can be computed using any of the following integration methods: Euler, Runge-Kutta, Implicit Euler or Implicit Midpoint. SODECL has previously been used to fit both ODE and SDE versions of an oculomotor control model to experimentally recorded timeseries~\cite{Avramidis2015,Avramidis2017}.

In the following sections, we describe the design principles of the SODECL library, its organisation and the numerical algorithms used. Moreover, we describe the experimental protocols that were implemented to measure the execution speed of SODECL for different compute devices ({\it i.e.} CPUs and GPUs), and to assess the numerical stability and accuracy of our SDE solver. As part of the speed test experiments, we compare the performance of SODECL on a high-end multi-core CPU with a MATLAB program that also calculates multiple SDE orbits in parallel using the same integration method ({\it i.e.} Euler--Maruyama). Lastly, we outline directions for future development of the library.

\section{The SODECL library}

\subsection{Design principles}

We chose to implement SODECL using OpenCL, because it does not pose significant constraints on the computer hardware used. As mentioned above, whilst a user must run CUDA executables on an NVIDIA GPU, OpenCL executables can run on both NVIDIA GPUs and AMD/Intel GPUs. Moreover, OpenCL executables can run on Intel and AMD CPUs. This gives OpenCL a key advantage over CUDA in terms of disseminating research methods and results, and promoting scientific collaborations.

We designed SODECL to be relatively portable, modular, and easy to extend. We tested our library on the following operating systems: Windows 10 64-bit; Ubuntu 18.04 64-bit; macOS Sierra (10.12). The SODECL library can easily be integrated into any C++ source code by including the main header of the library and linking it to an OpenCL library. Moreover, SODECL can be extended by adding OpenCL functions for other SDE integration methods ({\it e.g.} the Milstein method~\cite{Platen1999}). The source code is released under the MIT License and is under version control with \texttt{git} at \url{https://github.com/avramidis/sodecl}.

\subsubsection{The integration method}
SODECL uses the Euler-Maruyama method~\cite{Saito1996,Higham2001} to integrate the equations in (\ref{eq:SODEs}). The Euler-Maruyama approximation $\left\{\mathbf{X}^k=\left(X_1^k,\ldots,X_n^k\right): k\geq1\right\}$ to the true solution $\{\mathbf{X}(t): \mathbf{X}(t_0)=\mathbf{X}_0, t \geq t_0\}$ is defined by the recursion
\begin{equation}
\label{eq:EulerMurayama}
X^{k+1}_i  = X^k_i + f_i(t,\mathbf{X}^{k},\mathbf{p})\Delta t + \sum_{j=1}^{m}g_{ij}(t,\mathbf{X}^{k},\mathbf{p})\sqrt{\Delta t}N_j(0,1),
\end{equation}
where $\mathbf{X}^{1}=\mathbf{X}_0$, $\Delta t$ is the time step (so that $\mathbf{X}^{k}$ is the approximation to $\mathbf{X}((k-1)\Delta t)$ for $k>1$) and $N_j(0,1)$ are independent, normally distributed random variables with zero mean and unit variance.

\subsubsection{Noise generator}

To generate the noise variables $N_j(0,1)$ used in each step of ($\ref{eq:EulerMurayama}$), SODECL utilises the Random123 library~\cite{Salmon2011}. Random123 generates uniformly distributed random numbers, which are then converted to normally distributed random numbers using the Box-Muller algorithm~\cite{Box1958}{:}
\begin{align}\label{eq:BoxMuller}
\begin{array}{lcc}
r_1 =\sqrt{-2 \ln U_1} \cos(2 \pi U_2),\\
r_2 = \sqrt{-2 \ln U_1} \sin(2 \pi U_2).
\end{array}
\end{align}
Here, {$\{U_1,U_2\}$} are two independent, uniformly distributed random variables over [0,1] and {$\{r_1,r_2\}$} are two independent, normally distributed random variables with zero mean and unit variance. In the case where $j>2$, Eqs.~(\ref{eq:BoxMuller}) are called multiple times.

\subsubsection{OpenCL code generation}
SODECL creates and builds the OpenCL source code string at runtime. This allows the user to change the $f_i$ and $g_{ij}$ functions in (\ref{eq:SODEs}) without the need to compile the source code. Moreover, SODECL allows the implementation of wrappers -- {\it e.g.}, for MATLAB or Python, the use of a script/function that runs the SODECL executable, or the use of a MATLAB .mex file without it being recompiled. The procedure for formulating the OpenCL source code string is performed in the four steps outlined below.

\paragraph{Step 1}
Initially, SODECL appends the definitions of five named constants to the beginning of the empty OpenCL source code string (see Table~\ref{Table:NamedConstants}). The values of these constants depend on the general properties of the SDE system (numbers of equations, noise terms and parameters), the integrator's time step and the number of steps to be executed by the integrator in each OpenCL kernel call. These values are passed to the SODECL library by the user. The remainder of the OpenCL source code string is generated using the code located in three separate files, described in steps 2-4 below.


\begin{table}
	\caption{List of named constants used by the SODECL OpenCL functions.}
	\label{Table:NamedConstants}
	\begin{minipage}{\columnwidth}
		\begin{center}
			\begin{tabular}{cl}
				\toprule
				Named constant   & \multicolumn{1}{c}{Description}     \\
				\toprule
				\_numeq\_ 			& Number of equations in the SDE system	\\
				\_numnoi\_ 			& Number of noise variables in the SDE system	\\
				\_numpar\_			& Number of parameters in the SDE system	\\
				\_m\_dt\_			& SDE integrator time step in seconds			\\
				\_numsteps\_		& Number of SDE integrator steps per OpenCL kernel call \\
				\bottomrule
			\end{tabular}
		\end{center}
	\end{minipage}
\end{table}

\paragraph{Step 2}
The first file, \verb|integrator_caller.cl|, contains the kernel function that is called by the host to run on an OpenCL device. This function is defined in Figure~\ref{List:kernel}. The parameters of the kernel function are pointers that show the location of the arrays in the global memory of the device. These arrays contain the time points of the SDE system at which to approximate the solution, the values of the SDE system's dependent variables and the parameter combinations for which the different orbits are to be calculated. One additional array of counters is passed to the kernel for use by the Random123 library. The kernel generates the independent, normally distributed numbers required for the integration of the SDE system. Subsequently, the same kernel calls the function that integrates the SDE system for one time step (see step 3). The total number of calls is defined by the constant {\tt \_numsteps\_}. Having integrated the SDE system {\tt \_numsteps\_} times, the kernel copies the last integrated state values to the global memory of the OpenCL device, to be used in the next kernel step and accessed by the host.


\begin{figure}
	\begin{lstlisting}[language=C, frame=single]
	__kernel void integrator_caller(__global double *t,
	__global double *y,
	__global double *params_g,
	__global int *counter_g)
	\end{lstlisting}
	\caption{Definition of the SODECL kernel function. The function parameters are pointers showing the location of the arrays in the global memory of the OpenCL device. These arrays are the time points at which the solution to the SDE system is to be approximated (accessed by {\tt t}), the corresponding values of the dependent variables (accessed by {\tt y}) and the combinations of system parameters (accessed by {\tt params\_g}). One additional array of counters (accessed by {\tt counter\_g}) is passed to the kernel, for use by the Random123 library.}
	\label{List:kernel}
\end{figure}

\paragraph{Step 3}
The second file used in the creation of the OpenCL source is named \verb|stochastic_euler.cl|. It contains the function \verb|system_integrator| (defined in Figure~\ref{List:euler}), that implements the Euler-Muryama integration method. \verb|system_integrator| itself calls functions that calculate the deterministic and stochastic components of the SDE system (see step 4). Once these have been evaluated, \verb|stochastic_euler.cl| calculates the values of all dependent variables at the next time step.


\begin{figure}
	\begin{lstlisting}[language=C, frame=single]
	void system_integrator(double t, 
			       double y[_numeq_], 
			       double yout[_numeq_], 
			       double p[_numpar_], 
			       double noise[_numnoi])
	\end{lstlisting}
	\caption{Definition of the SODECL {\tt system\_integrator} function. The function parameters are the current time value of the SDE system ({\tt t}), the corresponding values of the dependent variables ({\tt y}), the parameter values ({\tt p}), the current noise values ({\tt noise}) and the new values of the dependent variables calculated by the Euler-Maruyama method ({\tt yout}).}
	\label{List:euler}
\end{figure}

\paragraph{Step 4}
The third file contains two functions called \verb|sode_system| and \verb|sode_system_stoch|, which calculate the drift and diffusion terms ${f_i}$ and $g_{ij}$ in the numerical scheme (\ref{eq:EulerMurayama}), respectively. Their definitions are shown in Figure~\ref{List:SDEsystem} and they have to be specified by the user (an example is shown in Figure~\ref{List:SDEKuramotoSDECL}).


\begin{figure}
	\begin{lstlisting}[language=C, frame=single]
	void sode_system(double t, 
			 double y[_numeq_], 
			 double yout[_numeq_], 
			 double p[_numpar_])
	
	void sode_system_stoch(double t, 
			       double y[_numeq_], 
			       double stoch[_numeq_], 
			       double p[_numpar_], 
			       double noise[_numnoi_])
	\end{lstlisting}
	\caption{Definition of the SODECL functions specifying the SDE system. The function parameters are the current time value ({\tt t}), the corresponding values of the dependent variables of the SDE system ({\tt y}) , the new values of the dependent variables of the deterministic component of the SDEs ({\tt yout}), the model parameters ({\tt p}), the noise values ({\tt noise}) and the new values of the stochastic components of the SDEs ({\tt stoch}).}
	\label{List:SDEsystem}
\end{figure}

\subsubsection{SDE integrator execution algorithm}

The logical flow of the SODECL integration execution algorithm is predominately a loop. During each loop iteration, the corresponding values of the dependent variables for all orbits are copied to the host device, and the copied values are saved in parallel to an array in preparation for the next kernel call. The OpenCL kernel is called to integrate the model for \verb|_numsteps_| steps of length \verb|_m_dt_|, which means that the output of the integration is saved to the array every (\verb|_numsteps_| $\times$ \verb|_m_dt_|) of integration time. This allows the data to be stored with a frequency of $1/$(\verb|_numsteps_| $\times$ \verb|_m_dt_|) Hz. Once the integration over the required time span is completed for all parameter combinations, the algorithm ends. The number of loop iterations, $k$, is calculated using the equation
\begin{align}\label{eq:KernelCallTimes}
k = t_{\rm total}/ (\Delta t \cdot s),
\end{align}
where $t_{\rm total}$ is the time span of the integration, $\Delta t$ is the integrator time step and $s$ is the number of steps the integrator performs in each OpenCL kernel call ({\it i.e.} the number of parallel operations, which could include multiple parameter combinations and/or initial conditions). The values of $t_{\rm total}$, $\Delta t$ and $s$ are supplied by the user.

The storing frequency of the results plays an important role in the speedup provided by the GPU compared to the CPU. This is due to the bottleneck caused by the transfer to and from the GPU memory, and the time required for storing the data to the host memory for further analysis. A low frequency (250-1000 Hz) avoids substantial data storage in the host memory. Also, the model size is an important factor, since it sets the number of model integrations that are to be saved on the computer's RAM. Moreover, a larger number of returned data points can increase the analysis time; a low number, however, may be insufficient for an accurate analysis \cite{Avramidis2017}.

\subsubsection{Python interface}

The Python package interface function for SODECL is defined in Figure~\ref{List:PythonInterface}, whilst Table~\ref{Table:PythonPackage} describes the arguments passed to the SODECL executable by the function. The Python library pybind11 (\url{https://github.com/pybind/pybind11}) was used to implement the interface. A MATLAB function interface is also provided with SODECL, although it should be noted that this is currently experimental.


\begin{table}
	\caption{Arguments of the SODECL Python interface.}
	\label{Table:PythonPackage}
	\begin{minipage}{\columnwidth}
		\begin{center}
			\begin{tabular}{cll}
			\toprule
			Argument 	 			& Variable 	& \multicolumn{1}{c}{Description} \\ 
			\toprule
			1   	 				& platform & OpenCL platform number \\ 
			2   	 				& device & OpenCL device number of the selected platform \\ 
			3   					& kernel & Path of the file with the OpenCL function defining the SDE system \\
			4   					& initx & Initial conditions for each orbit of the ODE system \\
			5   	 				& params & Parameter sets for the SDE system for all orbits \\
			6   	 				& integrator & SODE integrator \\
			7   					& orbits & Number of orbits to be calculated \\
			8   					& nequat & Number of equations of the SDE system \\
			9   					& nparams & Number of parameters of the SDE system \\ 
			10   					& nnoi & Number of noise processes \\ 
			11   					& dt & SDE solver time step \\
			12   					& tspan & Integration time span \\
			13   					& ksteps & Number of SDE integrator steps executed in each OpenCL kernel call \\  
			14 		 				& localgroupsize & OpenCL local group size \\
			\bottomrule
			\end{tabular}
		\end{center}
	\end{minipage}
\end{table}


\begin{figure}
	\begin{lstlisting}[language=Python, frame=single]
	def sodecl( platform, device, kernel,
		    initx, params, solver,
		    orbits, nequat, nparams,
		    nnoi, dt, tspan,
		    ksteps, localgroupsize )
	\end{lstlisting}
	\caption{Definition of the SODECL Python wrapper function. The function is used to execute the SODECL executable from within Python. Each argument is defined in Table \ref{Table:PythonPackage}.}
	\label{List:PythonInterface}
\end{figure}

\subsection{SODECL speed evaluation}

\subsubsection{Protocol}

To evaluate the execution speed of SODECL, we used a stochastic variant of the well-established Kuramoto model for describing the dynamics of a population of weakly coupled phase oscillators \cite{Daido92,Strogatz00,Acebron2005,Bick2011,Nakao16}. The Kuramoto model is commonly used to model synchronisation of biological oscillators~\cite{Acebron2005,Bick2011}, and is relatively straightforward to implement and customise ({\it e.g.} the oscillator population size can be easily varied). The SDE Kuramoto system we used here has the form
\begin{equation}
d\theta_i(t)=\omega_i+\frac{K}{N}\sum_{j=1}^{N}\sin(\theta_j(t)-\theta_i(t))+p_i dW_i(t),
\label{eq:StochKuramoto}
\end{equation}
where $\{\theta_1,\ldots,\theta_N: -\pi\leq\theta_i<\pi\}$ specify the phase of each oscillator, $\omega_i$ is the free-running frequency of the $i$th oscillator, $\{W_1(t),\ldots,W_N(t)\}$ are independent, scalar Wiener processes, $p_i$ controls the strength of the noise effect on the $i$th oscillator and $K$ is the strength of the mean-field coupling. The system parameters are thus $\mathbf{p}=(K,\omega_1,\ldots,\omega_N,p_1,\ldots,p_N)$.\footnote{In terms of Eq. (\ref{eq:SODEs}), $X_i=\theta_i$, $n=m=N$, $f_i=\omega_i+\frac{K}{N}\sum_{j=1}^{N}\sin(\theta_j-\theta_i)$ and $g_{ij}=p_i\delta_{ij}$.} The initial value of each phase variable $\theta_i$ was taken to be uniformly distributed over $[-\pi,\pi)$, the frequencies $\omega_i$ to be uniformly distributed over [0.01, 0.03] and the noise strengths $p_i$ to be uniformly distributed over [0.001, 0.003]. The following equations define the integration scheme for (\ref{eq:StochKuramoto}) using the Euler-Maruyama method ({\it cf.} (\ref{eq:EulerMurayama}) above):
\begin{align}\label{eq:KuramotoEulerMurayama}
\theta^{k+1}_i & = \theta^{k}_i + \left( \omega_i+\frac{K}{N}\sum_{j=1}^{N}\sin\left(\theta^{k}_j - \theta^{k}_i\right)\right)\Delta t + p_i \sqrt{\Delta t}N_i(0,1).
\end{align}
We set the coupling constant $K$ in \eqref{eq:StochKuramoto} to 1 and integrated the model for 400 s with a time step $\Delta t$ of $0.05$ s. Moreover, we used double-precision floating-point data type for better accuracy. The SODECL implementation of the stochastic Kuramoto model is shown in Figure~\ref{List:SDEKuramotoSDECL}.


\begin{figure}
	\begin{lstlisting}[language=C, frame=single]
	#define NOSC _numeq_
	
	void sode_system(double t, 
			 double y[_numeq_], 
			 double yout[_numeq_], 
			 double p[_numpar_])
	{	
	  for (int i = 0; i<NOSC; i++)
	  {
	    yout[i] = 0;
	    for (int j = 0; j<NOSC; j++)
	    {
	      yout[i] = yout[i] + (sin(y[j] - y[i]));
	    }
	    yout[i] = yout[i] * (p[0] / NOSC);
	    yout[i] = yout[i] + p[i+1];
	  }
	}
	
	void sode_system_stoch(double t, 
			       double y[_numeq_], 
			       double stoch[_numeq_], 
			       double p[_numpar_], 
			       double noise[_numnoi_])
	{
	  for (int i = 0; i < NOSC; i++)
	  {
	    stoch[i] = p[_numpar_ - _numnoi_ + i] * noise[i];
	  }
	}
	\end{lstlisting}
	\caption{SODECL implementation of the stochastic Kuramoto model (\ref{eq:StochKuramoto}). The first function {\tt sode\_system} is used to calculate the deterministic component of the model, whilst the second function {\tt sode\_system\_stoch} is used to calculate the stochastic part.}
	\label{List:SDEKuramotoSDECL}
\end{figure}

We compared the execution speed of SODECL when using different multi-core and many-core compute devices with the execution speed when using only one core of an HPC CPU. The compute devices used to evaluate SODECL performance are listed in Table~\ref{Table:Hardware}. The CPU with ID I4790K is a high-end desktop CPU, whereas the 2X6142 is an HPC node with two many-core CPUs. The W8100 and P100 are workstation/server GPUs, with high double-precision compute capabilities required for scientific computing. To gain a better understanding of how the SODECL runtime is affected by the size of the system, we varied the number of equations in the Kuramoto model ({\it i.e.} we varied the number of oscillators $N$ in the network). Moreover, we examined how the number of orbits $M$ being integrated affects the performance of SODECL on the different compute devices. Eight independent runs of the solver were executed for each choice of $N$ and $M$. 


\begin{table}
	\caption{Hardware used for the speed evaluation tests.}
	\label{Table:Hardware}
	\begin{minipage}{\columnwidth}
		\begin{center}
			\begin{tabular}{lll}
				\toprule
				\multicolumn{1}{c}{ID} & \multicolumn{1}{c}{Type} & \multicolumn{1}{c}{Model} \\				
				\toprule
				I4790K   	& CPU & Intel Core i7-4790K 			\\
				2X6142   	& CPU & 2 x Intel Xeon Gold 6142 		\\
				W8100   	& GPU & AMD Firepro W8100				\\
				P100        & GPU & NVIDIA Tesla P100-PCIE   		\\
				\bottomrule
			\end{tabular}
		\end{center}
	\end{minipage}
\end{table}

We also compared the execution speed of SODECL on a CPU against an equivalent parallel MATLAB implementation running on the same CPU. In order to further accelerate this implementation, described in Figure~\ref{List:SDEKuramotoMatlab}, we used the MATLAB Coder to convert MATLAB functions into C++ code. The parallelisation of MATLAB was performed automatically with OpenMP (Open Multi-Processing), an application programming interface for parallel applications that uses the cores of CPUs and coprocessors~\cite{Dagum1998}.


\begin{figure}
	\begin{lstlisting}[language=matlab, frame=single]
	function yout = kuramotoParallel(pop, nequat, x_y0, x_params, x_noise)
		
	  yout=zeros(pop,26);
		
	  parfor (p=1:pop,8)
	    y=kuramoto_system(nequat, x_y0(p,:), ...
	    x_params(p,:), x_noise(p,:));
	    yout(p,:)=y;
	  end
	end
		
	function yout = kuramoto_system(nequat, init, p, x_noise )
	  dt=5e-2;
	  sdt=sqrt(dt);
	  yout=zeros(1,26);
	  
	  for ii=1:25
	    for i=1:40
	      y=zeros(1,nequat);
	      for k=1:nequat
	        y(1,k)=0;
	        for j=1:nequat
	          y(1,k)=y(1,k)+sin(init(1,j)-init(1,k));
	        end
	        y(1,k)=p(1,k)+y(1,k)*p(1,nequat+1)/nequat;
	      end
	  
	      for j=1:nequat
	        init(1,j)=init(1,j)+y(1,j)*dt+x_noise(1,j)*sdt*randn(1);
	      end
	  
	      for k=1:nequat
	        y(1,k)=init(1,k);
	      end
	    end
	    yout(ii+1)=y(1,1);
	  end
	end
	\end{lstlisting}
	\caption{MATLAB function for calculating orbits of the stochastic Kuramoto model in parallel. The first function {\tt kuramotoParallel} calls the second function {\tt kuramoto\_system} -- which calculates one orbit of the model -- multiple times.} 
	\label{List:SDEKuramotoMatlab}
\end{figure}

\subsubsection{Results}

\begin{figure}
	\centering
	\includegraphics[scale=0.35]{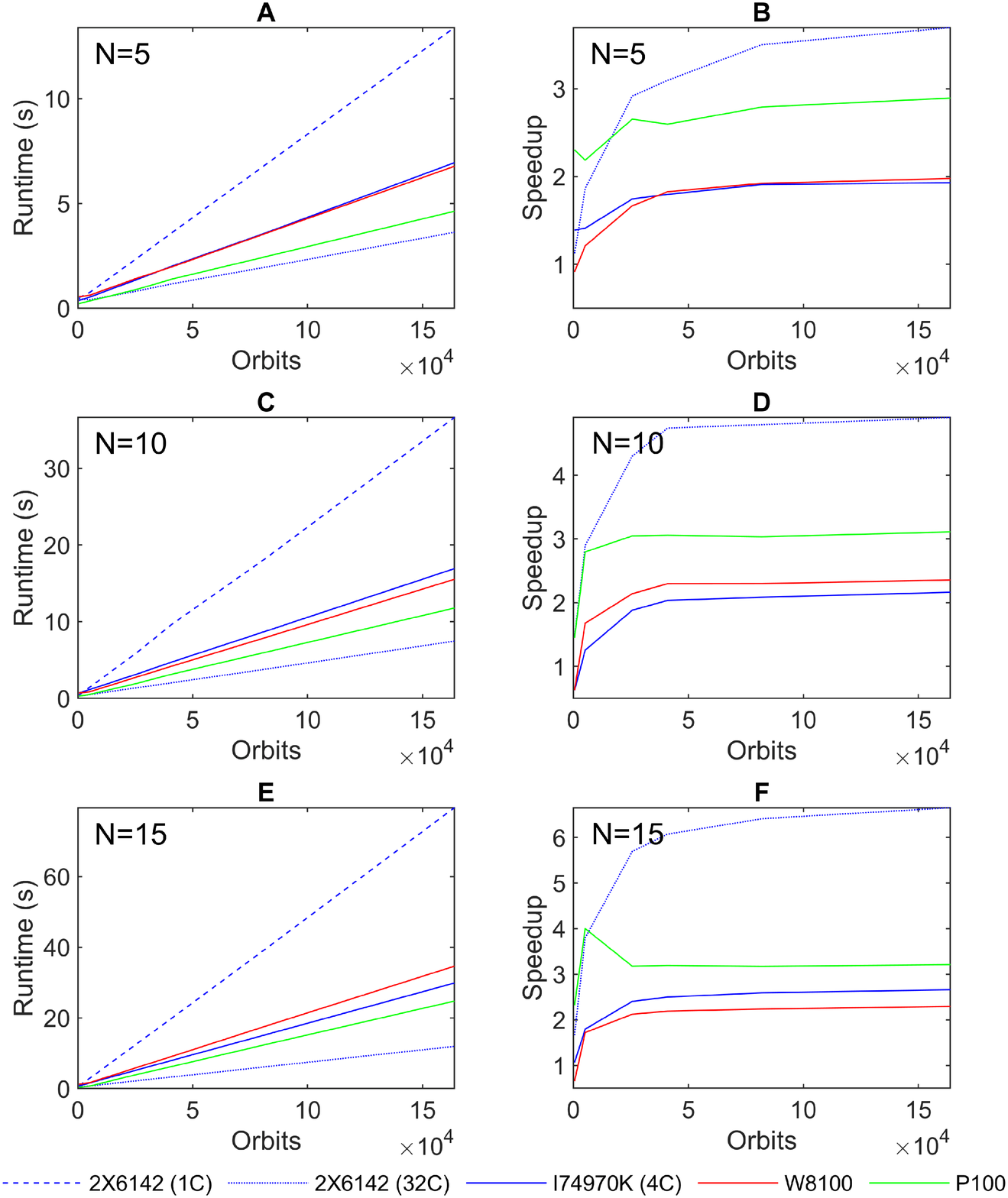}
	\caption{Runtimes in seconds (s) and speedups for the stochastic Kuramoto model as a function of the number of orbits for varying system sizes ($N$), when using different parallel computer hardware implementations. Plot A shows the runtimes for $N=5$, plot C for $N=10$ and plot E for $N=15$. The values are means from 8 runs. Plots B, D and F show the corresponding speedups for $N=5$, $N=10$ and $N=15$, respectively, relative to one logical processor of the 2X6142. The hardware used in each case is listed in Table~\ref{Table:Hardware}.}
	\label{Fig:results_kuramoto}
\end{figure}

The comparative execution speeds and speedups obtained using SODECL with different compute devices, numbers of orbits and system sizes are shown in Figure~\ref{Fig:results_kuramoto} (see Tables~\ref{Table:Runtimes5}-\ref{Table:speedup20} for the corresponding numerical values). For each compute device, the speedup was calculated relative to one core of the 2X6142 node ({\it i.e.} as the ratio of the 2X6142 single core runtime to the compute device runtime). The local group sizes used for each device and system size ($N$) are shown in Table~\ref{Table:lgbsize}. The local group size is the number of work items that will run in parallel and can communicate. In the case of embarrassingly parallel applications, although there is no communication, the group size can have an effect on the runtime. The values shown were found by doing exploratory tests, since -- to our knowledge -- there is no analytical formula for calculating the optimal local group size.


\begin{table}
	\caption{Local group sizes used for each compute device to generate the runtime results shown in Tables~\ref{Table:Runtimes5}-\ref{Table:speedup20} and plotted in Figure \ref{Fig:results_kuramoto}.}
	\label{Table:lgbsize}
	\begin{minipage}{\columnwidth}
		\begin{center}
			\begin{tabular}{ccccccc}
				\toprule
				Model size, $N$ & I4790K & 2X6142 & W8100 & P100 \\
				\toprule
				\es5  & 8 & 32 & 256 & 8 \\
				10 & 8 & 32 & \es16 & 8  \\
				15 & 8 & 32 & \es32 & 8  \\
				\bottomrule
			\end{tabular}
		\end{center}
	\end{minipage}
\end{table}

For the CPUs tested, the results indicate that when using all 32 cores of the 2X6142 node, SODECL is between $\simeq$1.12 and $\simeq$6.65 times faster than one core of the same node across different system sizes and orbit numbers. By contrast, when using all cores of the I4790K CPU and comparing its speed to one core of 2X6142, we observe speedups ranging from $\simeq$0.69 to $\simeq$2.66 as the system size is increased from 5 to 15 and the number of orbits is increased from 512 to 163840. The greater speedup obtained with 2X6142 is due to its larger number of cores and newer architecture.

For the GPUs tested, SODECL yields speedups in the range 0.6-4.5 compared to using one core of 2X6142, depending on the size of the Kuramoto model and the number of orbits. Larger numbers of model equations and orbits allow for greater use of GPU resources, which translates to improved speedup. When comparing the GPU runtimes to those obtained when all cores of 2X6142 are utilised, it can be seen that the GPUs are generally slower. This could, for example, be due to higher initialisation time overheads.

\begin{figure}
	\centering
	\includegraphics[scale=0.5]{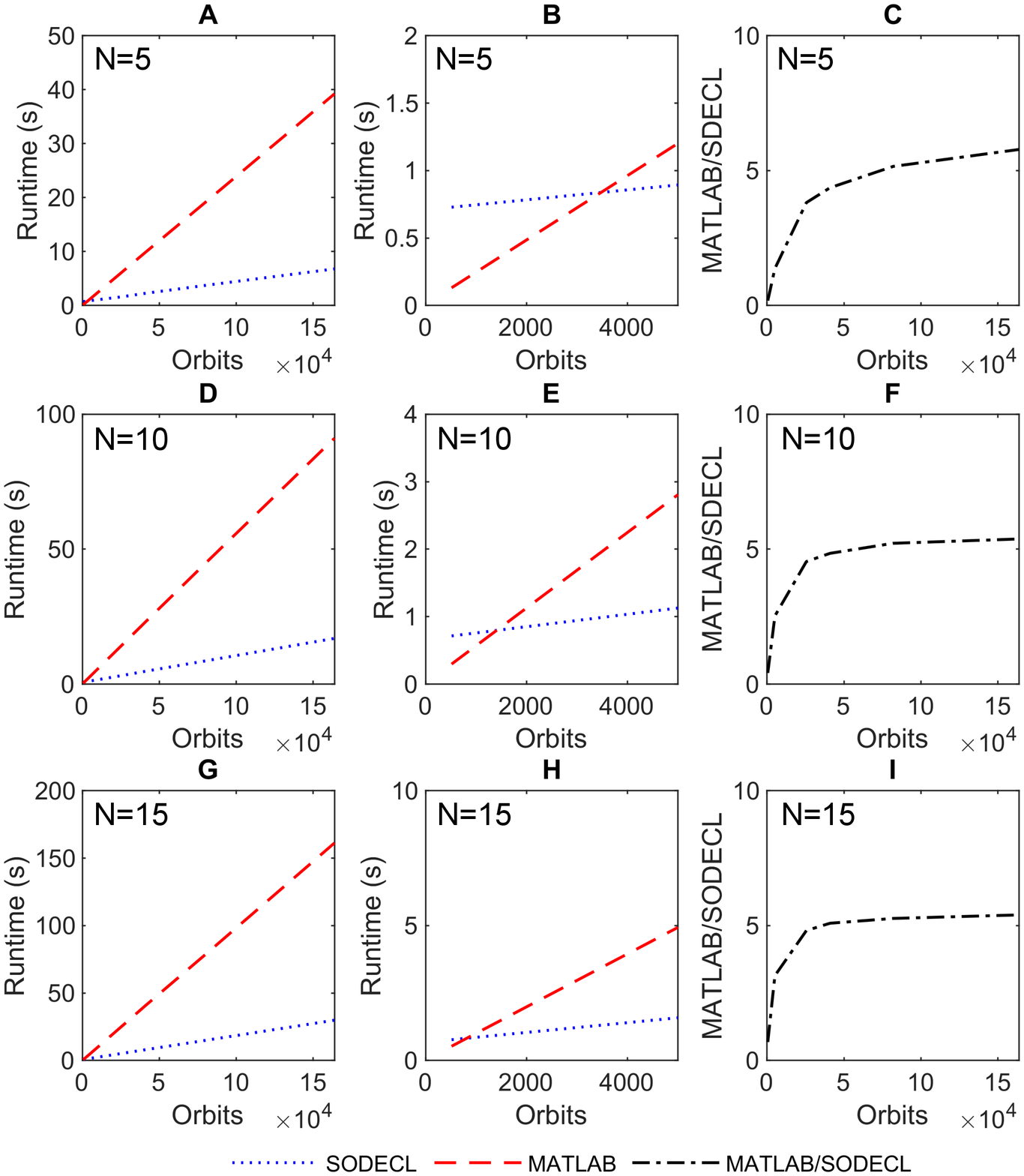}
	\caption{Runtime as a function of the number of orbits for the stochastic Kuramoto model for varying system sizes ($N$), using MATLAB and SODECL implementations on all logical processors of the I4790K CPU. Runtimes are given in seconds (s). Plots A \& B show results for $N=5$, D \& E for $N=10$ and G \& H for $N=15$. Plots B, E \& H show the runtimes obtained for the first 5000 orbits. Plots C, F \& I show the relative speeds of the MATLAB and SODECL implementations for $N=5,\,10$ and $15$, respectively.}
	\label{Fig:resultsmatlab}
\end{figure}

Finally, Figure~\ref{Fig:resultsmatlab} compares the runtimes obtained with the SODECL and MATLAB implementations of the Kuramoto model solver on the I4790K CPU  (the corresponding numerical values are given in Tables ~\ref{Table:MATSODECL5}-\ref{Table:MATSODECL15}). The MATLAB implementation is only faster than SODECL for the smallest system size and lowest number of orbits (see Figures ~\ref{Fig:resultsmatlab}B, E \& H). In all other cases, SODECL is up to 5.8 times faster.


\subsection{Accuracy and numerical stability tests}

\subsubsection{Protocol}

To assess the accuracy and numerical stability of SODECL, we generated a set of further simulations with $N=100$, $\omega_i\sim\mathcal{U}(0.2,0.4)$, $\phi_i(0)\sim\mathcal{U}(-\pi,\pi)$ and $p_i\sim\mathcal{U}(0.01,0.03)$, for two values $K_1=0.02$ and $K_2=0.2$ of the global coupling strength $K$. These values were chosen to lie either side of the Kuramoto transition that occurs in the deterministic system in the continuum limit ($N\to\infty$) at $K=K_c=\frac{2}{5\pi}\approx$ 0.1273. As $K$ is increased through this critical value, macroscopic mutual entrainment (MME) -- a collective synchronised rhythm with a common frequency -- is observed in the oscillator population \cite{Daido92,Strogatz00}. For both values of $K$, 64 independent integrations were carried out over 400 s (around 19 cycles of the mean period 20.94 s), using the following integration time steps: $\Delta t=\frac{2^{l-5}}{5}, 1\leq l\leq 5$. For each realisation, the degree of synchronisation of the simulated oscillator population was quantified using the complex order parameter \cite{Strogatz00,Nakao16}, defined below:
\begin{equation}
r(t)e^{i\Phi(t)} =\frac{1}{N}\sum_{j=1}^{N}e^{i\theta_j(t)}.
\label{eq:ordparam}
\end{equation}
In the above, the radius $r(t)$ measures the phase coherence and $\Phi(t)$ measures the collective phase, with $r$ values of 0 and 1 corresponding to complete desynchronisation and complete synchronisation of the population, respectively \cite{Strogatz00,Nakao16}. In the continuum limit of the deterministic model, increasing the coupling strength $K$ through the critical value $K_c$ causes the steady state coherence $\lim_{t \rightarrow \infty} r(t)$ to increase rapidly from 0 to 1 as the population becomes globally synchronised \cite{Strogatz00,Acebron2005,Nakao16}. For finite $N$, values of $K$ less than $K_c$ yield $\mathcal O(N^{-1/2})$ fluctuations in $r(t)$, while values of $K$ greater than $K_c$ result in  $r(t)$ saturating at a value $\lim_{t \rightarrow \infty} r(t)<1$, with $\mathcal O(N^{-1/2})$ fluctuations \cite{Strogatz00}. 

\subsubsection{Results}

\begin{figure}
	\centering
	\includegraphics[scale=0.5]{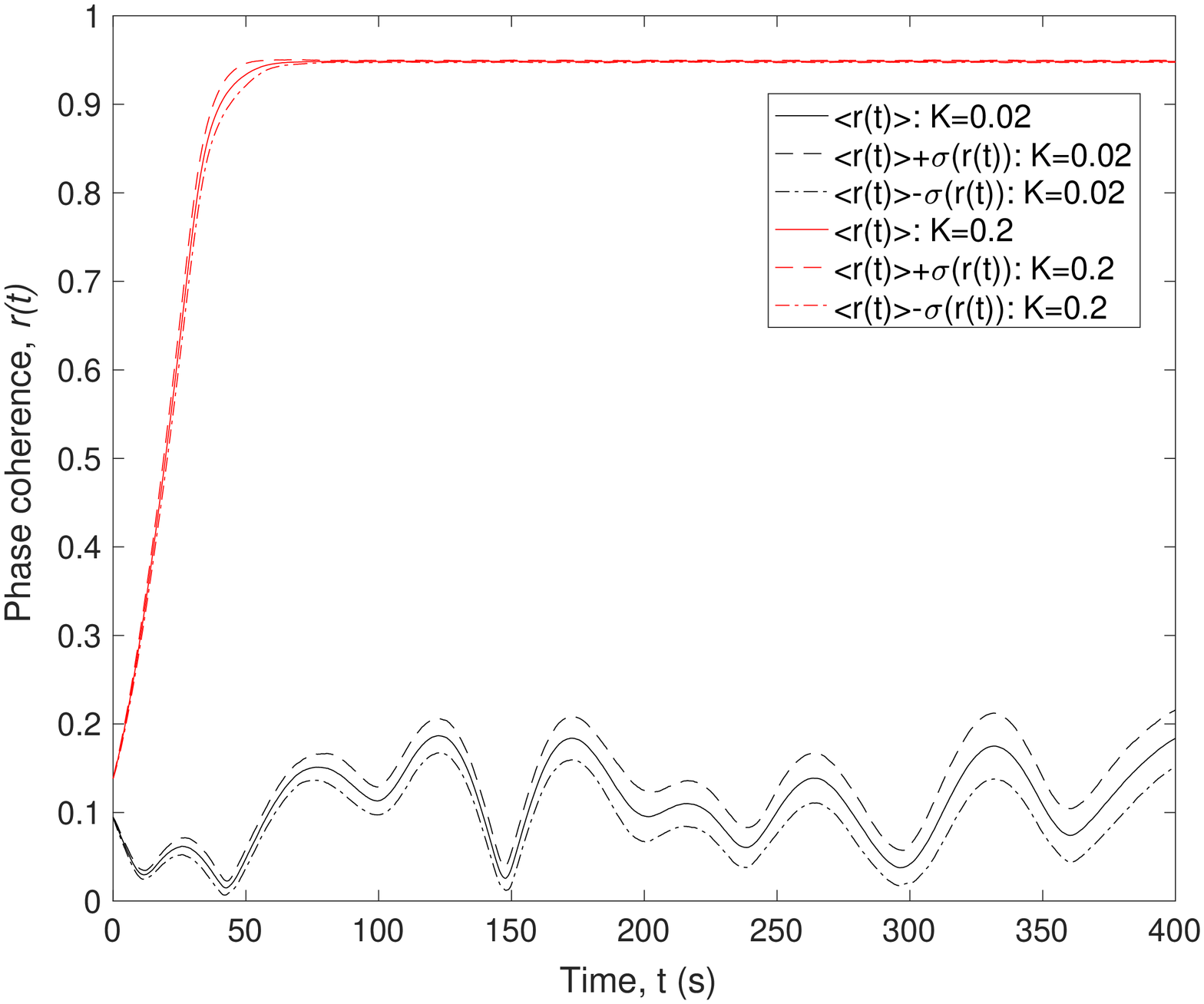}
	\caption{Temporal evolution of phase coherence $r(t)$ in the stochastic Kuramoto model (\ref{eq:StochKuramoto}) when $N=100$ for $K=0.02$ (black lines) and $K=0.2$ (red lines). $\langle r(t)\rangle$ and $\sigma \left(r(t)\right)$ denote the mean and standard deviation of $r(t)$, respectively, calculated from 64 independent realisations of (\ref{eq:KuramotoEulerMurayama}) with an integration timestep $\Delta t = 0.05$. Distributions of oscillator frequencies $\omega_i$, initial phases $\theta_i(0)$ and noise strengths $p_i$ were as described in the text. }
	\label{Fig:rordparam}
\end{figure}

\begin{figure}
	\centering
	\includegraphics[scale=0.525]{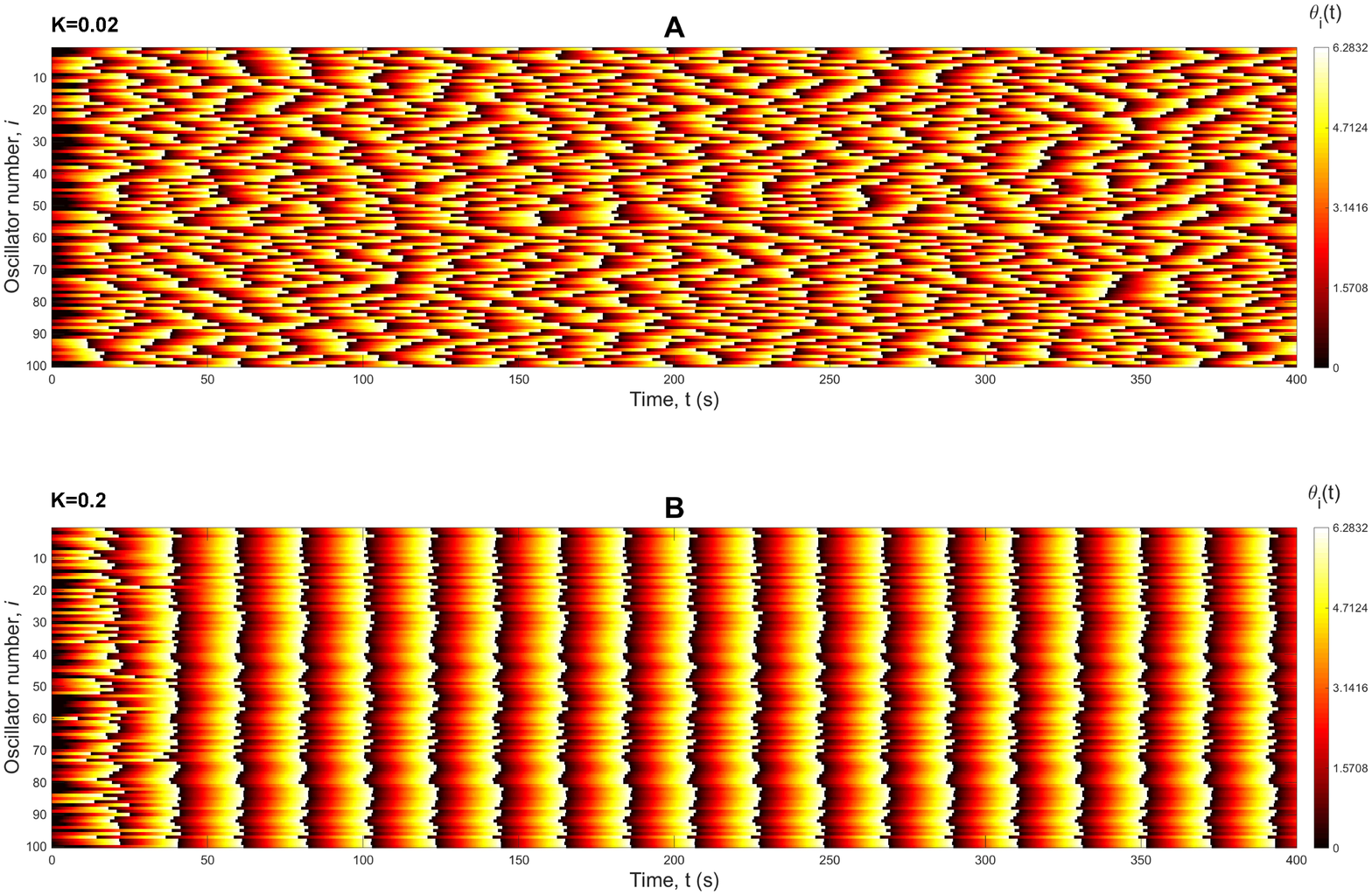}
	\caption{Temporal evolution of oscillator phases $\theta_i(t)$ in the stochastic Kuramoto model. Each kymograph corresponds to one of the realisations of (\ref{eq:KuramotoEulerMurayama}) used to compute the phase coherence statistics shown in Figure \ref{Fig:rordparam}.  Plots A and B shows the simulations for $K=0.02$ and $K=0.2$, respectively.}
	\label{Fig:phasesims}
\end{figure}

\begin{figure}
	\centering
	\includegraphics[scale=0.3]{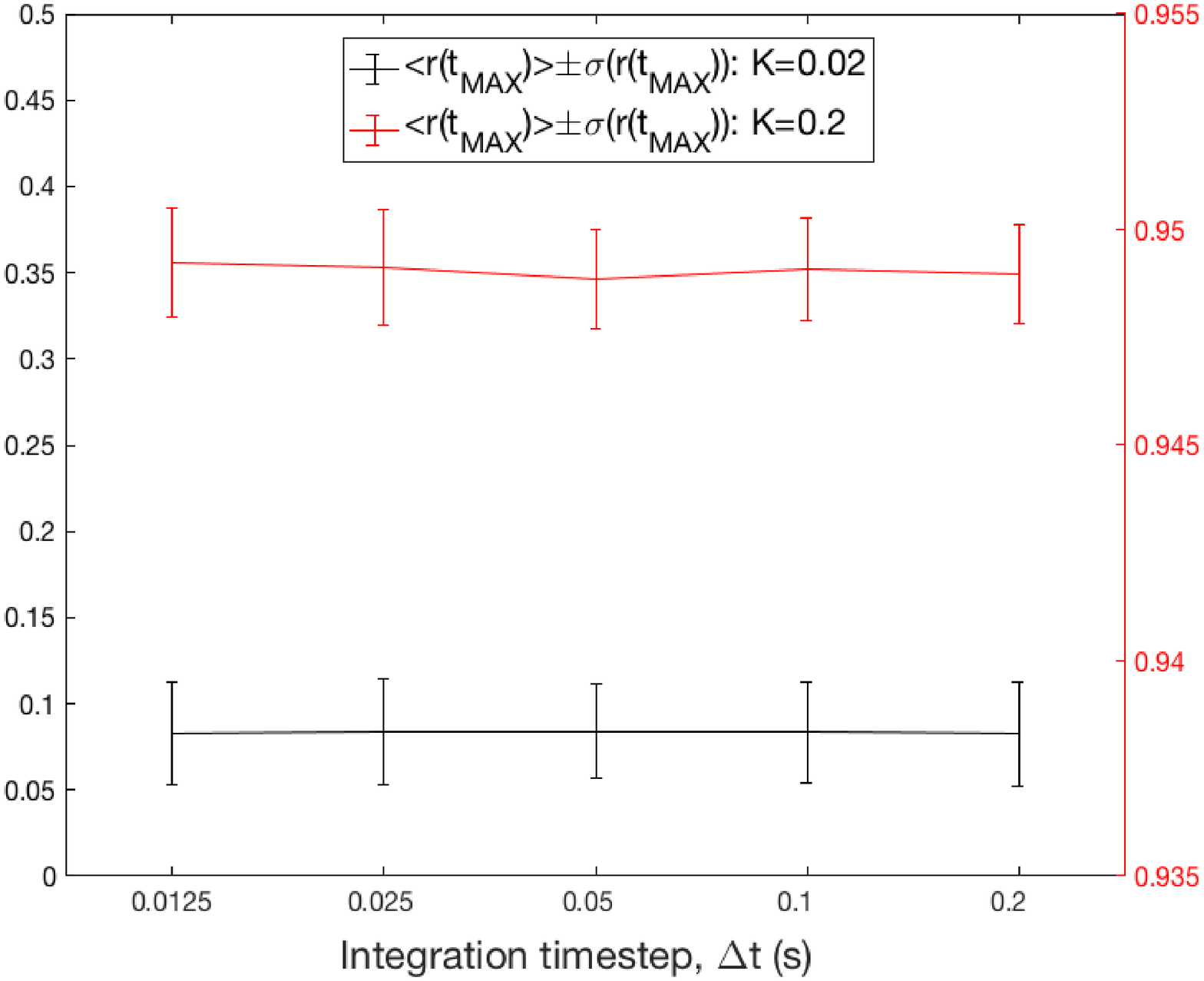}
	\caption{Dependence on the integration timestep $\Delta t$ of the phase coherence statistics $\langle r(t)\rangle$ and $\sigma \left(r(t)\right)$ at the end of the integration interval $t_{MAX}=400$. Black lines and axes show results for $K=0.02$; red lines and axes show results for $K=0.2$. In each case, $\langle r(t_{MAX})\rangle$ and $\sigma \left(r(t_{MAX})\right)$ were computed from 64 independent realisations of (\ref{eq:KuramotoEulerMurayama}) with $N=100$. Distributions of oscillator frequencies $\omega_i$, initial phases $\theta_i(0)$ and noise strengths $p_i$ were as described in the text. Note the logarithmic scale on the x-axis.}
	\label{Fig:stochscaling}
\end{figure}

Figure \ref{Fig:rordparam} plots how the mean phase coherence $\langle r(t)\rangle$ and the standard deviation of the phase coherence $\sigma \left(r(t)\right)$ vary with time for the two $K$ values when $\Delta t=0.05$. As expected, for $K=K_1<K_c$, $\langle r(t)\rangle$ exhibits small-amplitude oscillations close to 0, while for $K=K_2>K_c$, $\langle r(t)\rangle$ asymptotes to a value close to 1. Kymographs showing the temporal evolution of the population phases for a typical realisation generated with each $K$ value are plotted in Figure \ref{Fig:phasesims}. These are consistent with the phase coherence plots -- the oscillators are desynchronised throughout the integration interval for $K=K_1$, while for $K=K_2$, the oscillators synchronise after around 50 s and remain synchronised thereafter. Finally, Figure \ref{Fig:stochscaling} plots how the values of $\langle r(t)\rangle$ and $\sigma \left(r(t)\right)$ at the end of the integration interval $t_{MAX}=400$ vary with the integration timestep $\Delta t$. It can be seen that for both $K$ values, $\langle r(t_{MAX})\rangle$ and $\sigma \left(r(t_{MAX})\right)$ exhibit a weak dependence on $\Delta t$, suggesting that the integration algorithm is numerically stable \cite{Kloeden2011}.


\section{Conclusions}

We have described the design and performance of SODECL, an open source C++ library that uses OpenCL to calculate multiple orbits of a SDE system in an embarrassingly parallel way. The advantage of SODECL is that it can use the parallel capabilities of multi-core CPUs and GPUs to reduce computation time, thereby allowing large numbers of simulations to be performed simultaneously. This capability is critical when using population-based metaheuristics for data-fitting, such as genetic algorithms, as it facilitates the optimisation and analysis of SDE models in a practical time frame \cite{Adams13}. SODECL was designed to compile and run on Windows, Linux and macOS operating systems, to be user friendly, fast and fairly easy to extend.

Although execution speed was not the main priority in our design of SODECL, we showed that it was faster in almost all cases when integrating the stochastic Kuramoto system compared to a parallel MATLAB implementation. We also showed that when SODECL utilises a GPU or many-core CPUs, there is considerable speedup compared to a high-end desktop CPU. However, the speedup can be affected by the type and size of the SDE system being integrated. It would therefore be instructive to obtain performance benchmarks for other canonical stochastic models.

In our previous work~\cite{Avramidis2017}, SODECL achieved a speedup of $\simeq$20 on the W8100 GPU compared to all cores of the I4790K CPU. By contrast, the greatest speedup obtained with the same GPU against I4790K in this study was $\simeq$1.16. We believe that the difference in speedups between our previous and current studies is that in~\cite{Avramidis2017}, the SODECL OpenCL code could not be vectorised efficiently to facilitate better utilisation of the CPU, for the following reasons. Firstly, the system was a deterministic model with six heterogeneous equations ({\it i.e.} equations containing different functional forms). Similar equations, if vectorised, could run in parallel more efficiently. Secondly, two of these equations included a case mathematical function. Case functions are implemented with one or more {\tt if} statements that cause branch divergence. Most of the time, branch divergence cannot be predicted and a compiler cannot vectorise the code for branches in which each case implements a different mathematical operation. Thirdly, for the numerical integration of the model, the implicit, midpoint Euler method was used. This method includes a loop with a variable iteration number. All of the above factors hinder the compiler from efficiently vectorising the execution of the code on a parallel device.

There are a number of aspects of SODECL that we did not examine, which could provide interesting avenues for future work. Firstly, we only measured SODECL execution speed using double-precision, rather than single-precision, floating-point data type. Single-precision can be used if high accuracy is not necessary, and has the advantage that compute devices yield significantly better execution speeds when using lower precision. Secondly, we did not measure the OpenCL initialisation and finalisation time in SODECL. This would indicate whether different compute devices require more time for these tasks, perhaps explaining why the MATLAB implementation is comparable to SODECL for small numbers of orbits. Thirdly, we did not evaluate the effects of branch divergence on different compute devices using the Euler-Maruyama method. This potentially would have shown which devices are less affected. However, to assess the extent to which performance is affected by equations with many branches, a more extensive study would need to be carried out involving a broader range of SDE models. Lastly, we did not examine the memory requirements for different orbit numbers and model sizes. A follow-up study could examine the impact that the high memory requirements of larger models ({\it i.e.} more than 100 equations) could have on the performance of different compute devices.

We remark that the future development of SODECL would involve incorporating additional methods for numerically integrating SDE systems, such as the Milstein and Runge-Kutta schemes~\cite{Platen1999}. Here, we have implemented the Euler-Maruyama numerical integration scheme, which despite its simplicity, is still widely used, in part owing to the relative ease of implementation. We note, however, that the inefficiency of the Euler-Maruyama method compared to more sophisticated numerical schemes is offset by the significant acceleration conferred by parallelising the computation of trajectories across multiple CPU cores or GPU compute units. The incorporation of additional solvers would involve modifying the SODECL {\tt system\_integrator} function so as to implement the chosen schemes (see Fig.~\ref{List:euler}).

Also, additional optimisations could be explored for both GPUs and CPUs, such as using newer versions of OpenCL. This could potentially further accelerate SODECL on certain devices. Furthermore, we believe that splitting the integration algorithm over multiple OpenCL kernels could potentially increase the execution speed.

\section*{Acknowledgments}	
This work was financially supported by the Engineering and Physical Sciences Research Council (grant numbers EP/K040987/1, EP/N017846/1 and EP/N014391/1). Computational experiments were performed using resources provided by the Cambridge Service for Data Driven Discovery (CSD3, https://www.hpc.cam.ac.uk/high-performance-computing) and the University of Exeter High-Performance Computing (HPC) facility. CSD3 is operated by the University of Cambridge Research Computing Service (http://www.hpc.cam.ac.uk), funded by EPSRC Tier-2 capital grant EP/P020259/1, the STFC DiRAC HPC Facility (http://www.dirac.ac.uk) and the University of Cambridge. CSD3 and DiRAC are part of the UK National e-Infrastructure.

\bibliographystyle{unsrt}
\bibliography{Avramidisetal_arxiv}

\newpage

\section*{Supplementary Tables}
\setcounter{table}{0}
\renewcommand{\thetable}{S\arabic{table}}


\begin{table}[H]
	\caption{Runtimes in seconds (s) for the stochastic Kuramoto model as a function of the number of orbits for system size $N=5$, when using different parallel computer hardware implementations. The hardware used in each case is shown in Table~\ref{Table:Hardware}.}
	\label{Table:Runtimes5}
	\begin{minipage}{\columnwidth}
		\begin{center}
			\begin{tabular}{lccccccc}
				\toprule								
				\multicolumn{1}{c}{\es Orbits}	& \multicolumn{1}{c}{\es 2X6142 (1C)} & \multicolumn{1}{c}{\es 2X6142 (32C)} & \multicolumn{1}{c}{\es I4790K (4C)} &	\multicolumn{1}{c}{W8100} & \multicolumn{1}{c}{P100} \\
				\toprule
				\es\es\es512	& \es0.5044  & \es0.449  & \es0.728	& 0.552	& 0.201 \\
				\es\es5120 		& \es0.7463  & \es0.401  & \es0.898	& 0.617	& 0.363 \\
				\es25600 		& \es2.3734  & \es0.814  & \es1.612	& 1.424	& 0.935 \\
				\es40960 		& \es3.6019  & \es1.163  & \es2.254	& 1.973	& 1.400 \\
				\es81920 		& \es6.8686  & \es1.961  & \es3.800	& 3.573	& 2.512 \\
				163840 			& \es13.390  & \es3.623  & \es6.779	& 6.772	& 4.719 \\
				\bottomrule
			\end{tabular}
		\end{center}
	\end{minipage}
\end{table}


\begin{table}[H]
	\caption{Runtimes in second (s) for the stochastic Kuramoto model as a function of the number of orbits for system size $N=10$, when using different parallel computer hardware implementations. The hardware used in each case is shown in Table~\ref{Table:Hardware}.}
	\label{Table:Runtimes10}
	\begin{minipage}{\columnwidth}
		\begin{center}
			\begin{tabular}{lccccccc}
				\toprule
				\multicolumn{1}{c}{\es Orbits}	& \multicolumn{1}{c}{\es 2X6142 (1C)} & \multicolumn{1}{c}{\es 2X6142 (32C)} & \multicolumn{1}{c}{\es I4790K (4C)} &	\multicolumn{1}{c}{W8100} & \multicolumn{1}{c}{P100} \\
				\toprule				
				\es\es\es512	& \es0.441	&	\es0.301 &	\es0.714 &	\es0.708 	& 0.280 \\
				\es\es5120		& \es1.444	&	\es0.499 &	\es1.138 &	\es0.864	& 0.513 \\
				\es25600		& \es5.956	&	\es1.385 &	\es3.160 &	\es2.784	& 1.714 \\
				\es40960		& \es9.676	&	\es2.042 &	\es4.746 &	\es4.211	& 2.566 \\
				\es81920		& 18.308	&	\es3.821 &	\es8.784 &	\es7.981	& 4.828 \\
				163840			& 36.641	&	\es7.471 &	16.939   &	15.589		& 9.315 \\
				\bottomrule
			\end{tabular}
		\end{center}
	\end{minipage}
\end{table}


\begin{table}[H]
	\caption{Runtimes in second (s) for the stochastic Kuramoto model as a function of the number of orbits for system size $N=15$, when using different parallel computer hardware implementations. The hardware used in each case is shown in Table~\ref{Table:Hardware}.}
	\label{Table:Runtimes15}
	\begin{minipage}{\columnwidth}
		\begin{center}
			\begin{tabular}{lccccccc}
				\toprule
				\multicolumn{1}{c}{\es Orbits}	& \multicolumn{1}{c}{\es 2X6142 (1C)} & \multicolumn{1}{c}{\es 2X6142 (32C)} & \multicolumn{1}{c}{\es I4790K (4C)} &	\multicolumn{1}{c}{W8100} & \multicolumn{1}{c}{P100} \\
				\toprule				
				\es\es\es512	& \es0.8247	& \es0.495	& \es0.772	& \es1.271	& \es0.323 \\
				\es\es5120		& \es2.8846	& \es0.759	& \es1.605	& \es1.679	& \es0.672 \\
				\es25600		& 12.6576	& \es2.223	& \es5.256	& \es5.968	& \es2.940 \\
				\es40960		& 19.9707	& \es3.288	& \es7.987	& \es9.128	& \es4.566 \\
				\es81920		& 39.6412	& \es6.185	& 15.279	& 17.736	& \es9.056 \\
				163840			& 79.5124	& \es11.953	& 29.882	& 34.700	& 17.656   \\
				\bottomrule
			\end{tabular}
		\end{center}
	\end{minipage}
\end{table}



\begin{table}
	\caption{Speedups obtained for the stochastic Kuramoto model as a function of the number of orbits for system size $N=5$, when using each of the compute devices listed in Table~\ref{Table:Hardware}. Speedups are calculated as the runtime for one core of the Intel Xeon 6142 CPU divided by the runtime of the compute device. In the case of the CPUs (2X6142 and 4790K), all cores were used.}
	\label{Table:speedup3}
	\begin{minipage}{\columnwidth}
		\begin{center}
			\begin{tabular}{rcccccc}
				\toprule
				\multicolumn{1}{c}{\es Orbits} & \multicolumn{1}{c}{\es 2X6142} & \multicolumn{1}{c}{\es I4790K} &	\multicolumn{1}{c}{W8100} & \multicolumn{1}{c}{P100} \\
				\toprule				
				512	&   1.122 & 0.692 & 0.914 &	2.498\\
				5120& 	1.861 & 0.831 & 1.209 &	2.051\\
				25600& 	2.915 & 1.471 & 1.666 & 2.537\\
				40960& 	3.095 & 1.597 & 1.826 & 2.572\\
				81920& 	3.501 & 1.807 & 1.922 & 2.733\\
				163840& 3.696 &	1.975 & 1.977 & 2.837\\
				\bottomrule
			\end{tabular}
		\end{center}
	\end{minipage}
\end{table}


\begin{table}
	\caption{Speedups obtained for the stochastic Kuramoto model as a function of the number of orbits for system size $N=10$, when using each of the compute devices listed in Table~\ref{Table:Hardware}. Speedups are calculated as the runtime for one core of the Intel Xeon 6142 CPU divided by the runtime of the compute device. In the case of the CPUs (2X6142 and 4790K), all cores were used.}
	\label{Table:speedup10}
	\begin{minipage}{\columnwidth}
		\begin{center}
			\begin{tabular}{rcccccc}
				\toprule
				\multicolumn{1}{c}{\es Orbits} & \multicolumn{1}{c}{\es 2X6142} & \multicolumn{1}{c}{\es I4I90K} &	\multicolumn{1}{c}{W8100} & \multicolumn{1}{c}{P100} \\
				\toprule
				512	  & 1.462	&0.617	&0.622	&1.574\\
				5120 &2.895	&1.269	&1.671	&2.814\\
				25600 &4.301	&1.884	&2.139	&3.474\\
				40960 &4.738	&2.039	&2.298	&3.771\\
				81920 &4.791	&2.084	&2.294	&3.792\\
				163840 &4.905	&2.163	&2.350	&3.934\\
				\bottomrule
			\end{tabular}
		\end{center}
	\end{minipage}
\end{table}


\begin{table}
	\caption{Speedups obtained for the stochastic Kuramoto model as a function of the number of orbits for system size $N=15$, when using each of the compute devices listed in Table~\ref{Table:Hardware}. Speedups are calculated as the runtime for one core of the Intel Xeon 6142 CPU divided by the runtime of the compute device. In the case of the CPUs (2X6142 and 4790K), all cores were used.}
	\label{Table:speedup20}
	\begin{minipage}{\columnwidth}
		\begin{center}
			\begin{tabular}{rcccccc}
				\toprule
				\multicolumn{1}{c}{\es Orbits} & \multicolumn{1}{c}{\es 2X6142} & \multicolumn{1}{c}{\es I4790K} &	\multicolumn{1}{c}{W8100} & \multicolumn{1}{c}{P100} \\
				\toprule
				\es\es512	&   1.664	&	1.068	&	0.649	&	2.552\\
				\es5120		&	3.796	&	1.796	&	1.718	&	4.289\\
				\es25600	&	5.692	&	2.408	&	2.121	&	4.305\\
				\es40960	&	6.073	&	2.500	&	2.188	&	4.373\\
				\es81920	&	6.409	&	2.594	&	2.235	&	4.377\\
				163840		&	6.652	&	2.661	&	2.291	&	4.503\\				
				\bottomrule
			\end{tabular}
		\end{center}
	\end{minipage}
\end{table}


\begin{table}
	\caption{Runtimes for the stochastic Kuramoto model as a function of the number of orbits for system size $N=5$, using SODECL and parallel MATLAB implementations on all cores of the i7-4970K CPU. Runtimes are given in seconds (s). The last column (MATLAB/SODECL) shows the ratio of the runtimes of the MATLAB and SODECL implementations (the speedup). Speedup values greater than one indicate that SODECL was faster.}
	\label{Table:MATSODECL5}
	\begin{minipage}{\columnwidth}
		\begin{center}
			\begin{tabular}{rccc}
				\toprule
				\multicolumn{1}{c}{\es Orbits} & \multicolumn{1}{c}{SODECL} & \multicolumn{1}{c}{\es MATLAB} & \multicolumn{1}{c}{MATLAB/SODECL} \\
				\toprule				
				\es\es512	& \es0.728	& \es0.131	& \es0.180\\
				\es5120		& \es0.898	& \es1.229	& \es1.369\\
				\es25600	& \es1.612	& \es6.149	& \es3.813\\
				\es40960	& \es2.254	& \es9.841	& \es4.364\\
				\es81920	& \es3.800	& 19.588	& \es5.154\\
				163840		& \es6.779	& 39.213	& \es5.784\\
				\bottomrule
			\end{tabular}
		\end{center}
	\end{minipage}
\end{table}


\begin{table}
	\caption{Runtimes for the stochastic Kuramoto model as a function of the number of orbits for system size $N=10$, using SODECL and parallel MATLAB implementations on all cores of the i7-4970K CPU. Runtimes are given in seconds (s). The last column (MATLAB/SODECL) shows the ratio of the runtimes of the MATLAB and SODECL implementations (the speedup). Speedup values greater than one indicate that SODECL was faster.}
	\label{Table:MATSODECL10}
	\begin{minipage}{\columnwidth}
		\begin{center}
			\begin{tabular}{rccc}
				\toprule
				\multicolumn{1}{c}{\es Orbits} & \multicolumn{1}{c}{SODECL} & \multicolumn{1}{c}{\es MATLAB} & \multicolumn{1}{c}{MATLAB/SODECL} \\
				\toprule
				\es\es512	&	\es0.714	&	\es0.2966	&	0.415\\
				\es5120		&	\es1.1383	&	\es2.8749	&	2.525\\
				\es25600	&	\es3.1609	&	14.3615		&	4.543\\
				\es40960	&	\es4.7469	&	23.0054		&	4.846\\
				\es81920	&	\es8.784	&	45.8372		&	5.218\\
				163840		&	16.939		&	91.1109		&	5.378\\				
				\bottomrule
			\end{tabular}
		\end{center}
	\end{minipage}
\end{table}


\begin{table}
	\caption{Runtimes for the stochastic Kuramoto model as a function of the number of orbits for system size $N=15$, using SODECL and parallel MATLAB implementations on all cores of the i7-4970K CPU. Runtimes are given in seconds (s). The last column (MATLAB/SODECL) shows the ratio of the runtimes of the MATLAB and SODECL implementations (the speedup). Speedup values greater than one indicate that SODECL was faster.}
	\label{Table:MATSODECL15}
	\begin{minipage}{\columnwidth}
		\begin{center}
			\begin{tabular}{rccc}
				\toprule
				\multicolumn{1}{c}{\es Orbits} & \multicolumn{1}{c}{SODECL} & \multicolumn{1}{c}{\es MATLAB} & \multicolumn{1}{c}{MATLAB/SODECL} \\
				\toprule
				\es\es512	&	\es0.772	&	0.529	&	0.6861\\
				\es5120		&	\es1.606	&	5.050	&	3.144\\
				\es25600	&	\es5.256	&	25.322	&	4.817\\
				\es40960	&	\es7.987	&	40.642	&	5.087\\
				\es81920	&	15.279		&	80.467	&	5.266\\
				163840		&	29.882		&	161.250	&	5.396\\
				\bottomrule
			\end{tabular}
		\end{center}
	\end{minipage}
\end{table}

\end{document}